\documentclass[11pt]{article}
\usepackage{latexsym}
\usepackage{amssymb}
\usepackage[dvips]{graphicx}
\usepackage{epsfig}
\usepackage{rotating}
\usepackage{amsfonts}
\usepackage{amsmath}
\begin{document}

\begin{titlepage}
\begin{flushright}
{\  }\\%CP3-12-34\\
{\ }\\%ICMPA-MPA/026/2012\\
\end{flushright}

\vspace{10pt}

\begin{center}

{\Large\bf Time Dependent Conserved Charges and their Gauging}\\

\vspace{5pt}

{\Large\bf -- A Modest Case Study in Shared Memory of Victor --}\\

\vspace{40pt}

Jan Govaerts$^{a,b}$

\vspace{10pt}

$^{a}${\sl Centre for Cosmology, Particle Physics and Phenomenology (CP3),\\
Institut de Recherche en Math\'ematique et Physique (IRMP),\\
Universit\'e catholique de Louvain (UCLouvain),\\
2, Chemin du Cyclotron, B-1348 Louvain-la-Neuve, Belgium}\\
E-mail: {\em Jan.Govaerts@uclouvain.be}

\vspace{10pt}

$^{b}${\sl International Chair in Mathematical Physics and Applications (ICMPA--UNESCO Chair),\\
University of Abomey--Calavi, 072 B. P. 50, Cotonou, Republic of Benin}

\vspace{10pt}

%\today

\vspace{10pt}

\begin{abstract}
\noindent
There exist instances of dynamical systems possessing symmetry transformations of which the conserved Noether charges
generating these symmetries feature an explicit time dependence in their functional representation over phase space.
The generators of such symmetries certainly do not commute with the Hamiltonian, and yet these charges are conserved
observables for the classical and quantised dynamics. Furthermore within the Hamiltonian formalism and in the case of global
symmetries such charges may be gauged to allow for arbitrary time dependent symmetry transformations, simply
by extending the Hamiltonian to include the Noether charges as first-class constraints. An explicit illustration of these
issues is presented in a simple and most familiar model that applies also to the constant gravitational force.
This note  draws its primary motivation from the quest towards a theory for quantum gravity, in wanting to understand better the tension existing between
the local Equivalence Principle of the gravitational interaction and the fundamental principles of Quantum Mechanics
by considering the formulation of quantum systems relative to reference frames that are inertial or noninertial, and thus
accelerated relative to one another through arbitrary time dependent spatial translations.
\end{abstract}

\end{center}

\vspace{40pt}

\begin{center}
Published in the\\
Proceedings volume of the Victor Villanueva Memorial Workshop,\\
Morelia, Mexico, December 6, 2013,\\
Adnan Bashir and Christian Schubert (eds.),\\
{\sl J. Phys. Conf. Ser.} 1208 (2019) no. 1, 012011.
\end{center}

\end{titlepage}

\setcounter{footnote}{0}

\section{Introduction}
\label{Intro}

Whether for classical or quantum physical systems, it may seem to be a widely held belief that any conserved quantity, say $\hat{C}$,
related to a symmetry is characterised by an observable which commutes with the Hamiltonian operator, $[\hat{C},\hat{H}]=0$.
However this is not necessarily so, provided of course that the conserved charge possesses an explicit
functional time dependence---whether as a function over classical phase space, or as an operator acting on the Hilbert space of quantum states---which is
such that in the Heisenberg picture of quantum mechanics its Schr\"odinger equation of motion reads
(or for the corresponding Poisson brackets of the classical dynamics),
\begin{equation}
i\hbar\frac{d\hat{C}(t)}{dt}=i\hbar\frac{\partial\hat{C}(t)}{\partial t}\,+\,\left[\hat{C}(t),\hat{H}(t)\right]=0,\quad {\rm namely},\quad
\left[\hat{C}(t),\hat{H}(t)\right]=-i\hbar\frac{\partial\hat{C}(t)}{\partial t}.
\end{equation}
Even though not widely known, such instances of time dependent conserved quantities arise even for extremely simple
and common systems, which are such that the Lagrange function accounting for the system's dynamics is not left invariant
under the symmetry transformation associated to the conserved charge but rather, is invariant up to a total time derivative
term or surface term of an explicitly time dependent quantity, ensuring thereby the invariance of the system's Euler-Lagrange equations
of motion under the symmetry transformation.

Furthermore given the possibility of time\footnote{Or space-time dependent ones in the case of relativistic local field theories.}
dependent conserved charges associated to an ensemble of symmetry transformations,
one may also wonder whether it is possible to gauge such symmetries of which not only the parameters would be (space-)time
dependent but their generators as well.

The present note explores these issues in the simplest such instance of a time dependent conserved charge,
namely that of a system of a nonrelativistic massive particle subjected to a constant external force in some Euclidean space.
Indeed Newton's equations of that system are invariant under spatial translations, and yet, the corresponding
conserved generator is explicitly time dependent. Physical cases in point are that of the homogeneous gravitational
force leading to a constant acceleration $\vec{g}$, or of a static and homogeneous electric field\footnote{The dual case
of a static and homogeneous magnetic field is also invariant under spatial translations while in that case as well
the Lagrange function is invariant under these symmetries only up to a surface term. Spatial translations are then generated
by conserved charges that are then space dependent (rather than time dependent). This situation corresponds
to the celebrated Landau problem with its Bopp shift operators generating translations in the plane perpendicular to the magnetic field.}.

Incidentally, this author came to the exploration of such simple questions by pondering on the tension that exists between the local Equivalence Principle
of General Relativity and the fundamental principles of Quantum Mechanics with the spatially nonlocal characteristics of its dynamics.
In particular one may ask how to construct the dynamics of quantum systems in noninertial frames, {\it i.e.}, accelerated frames, and finally make the quantum dynamics
even frame independent, namely invariant under arbitrary space (or space-time) diffeomorphisms, thus in particular beginning
by gauging spatial translations in the simplest context possible. That the case of a motion of constant acceleration---as is the case
for gravity on the Earth's surface---immediately provides the simplest such illustration is rather reassuring.

The present modest note is offered in memory of our colleague and friend, Victor Villanueva. Ever since when he started onwards his scientific
path, Victor has remained interested in the fundamental issues bridging between quantum dynamics, symmetries and geometry, and their local realisations
through gauge transformations\cite{VV1,VV2,VV3,VV4,VV5}. As the true scientific adventurer in spirit that he was, I believe that he too would have liked to explore what
may well lie hidden beyond the modest starting paths outlined above. Possibly Victor would have been intrigued as much as I am by questions such as those
raised above, and then to be considered within the wider context of the open quest for a theory of quantum gravity. The algebraic structures of
Hilbert spaces and noncommutative geometries, and their operators and symmetries, may well hold the key to that ultimate goal of
fundamental theoretical physics at the frontiers of the XXI$^{\rm st}$ century.

\section{A Classical Little Story}
\label{Sect2}

To keep to the simplest nonrelativistic setting conceivable, let us consider a single degree of freedom system of Euclidean coordinate $x(t)\in\mathbb{R}$ relative to some
inertial frame, of mass $m$, subjected to a constant and static external force $F$---henceforth let us assume $F>0$, without loss of generality (by changing the
orientation of $x$ if need be)---of which the dynamics is accounted for by the following choice of action, Lagrange function, and potential energy,
\begin{equation}
S_0[x]=\int\,dt\,L_0(x(t),\dot{x}(t)),\qquad
L_0(x,\dot{x})=\frac{1}{2}m\dot{x}^2\,+\,xF,\qquad
V(x)=-xF.
\label{eq:S0}
\end{equation}
Equivalently in its Hamiltonian formulation the same dynamics is accounted for by the first-order action and Lagrange function, and Hamiltonian,
\begin{equation}
S_1[x,p]=\int\,dt\,L_1,\qquad
L_1=\dot{x}p\,-\,H_0(x,p),\qquad
H_0(x,p)=\frac{1}{2m}p^2\,+\,V(x)=\frac{1}{2m}p^2\,-\,xF,
\end{equation}
such that $(x,p)\in\mathbb{R}^2$ are canonically conjugate phase space variables with canonical Poisson bracket, $\{x,p\}=1$, and of course
the relation $p=\partial L_0/\partial\dot{x}=m\dot{x}$.

\subsection{Classical dynamics}
\label{Sect2.1}

The equations of motion of the system are well known, and read of course, whether in the Lagrangian or Hamiltonian formulations,
\begin{equation}
m\ddot{x}=F,\qquad
\dot{x}=\frac{1}{m}p,\qquad
\dot{p}=F.
\end{equation}
Given the initial values at $t=0$ as integration constants, $x_0=x(0)$, $v_0=\dot{x}(0)$, and $p_0=p(0)=mv_0$, one has
for the classical solutions,
\begin{equation}
x(t)=\frac{1}{2}\frac{F}{m}t^2\,+\,v_0 t \,+\, x_0,\qquad
\dot{x}(t)=\frac{F}{m} t + v_0,\qquad
p(t)=F t + p_0.
\end{equation}
A direct substitution for the energy of any such solution then finds,
\begin{equation}
E=H_0(x,p)=\frac{1}{2m}p^2+V(x)=\frac{1}{2m}p^2(t)-x(t)F=\frac{1}{2m}p^2_0-x_0F=\frac{1}{2}mv^2_0-x_0F,
\end{equation}
which is thus a time independent conserved quantity, while the following combination
defines an explicitly time dependent observable over phase space and yet a conserved quantity as well,
\begin{equation}
T(p,t)=p\,-\,t F=p(t)\,-\,t F=p_0=T.
\end{equation}
Note that by choosing appropriately the values for $x_0$ and $p_0$, there always exists a classical solution however
arbitrary the choice for these two conserved quantities in their respective ranges,
$-\infty<E<+\infty$ and $-\infty< T<+\infty$. In particular, given any value for $T\in\mathbb{R}$ there exists a solution for whatever
value of the energy, $E\in\mathbb{R}$, and vice-versa, given any value for $E\in\mathbb{R}$ there exists a solution for whatever value of $T\in\mathbb{R}$.
In other words there exists an infinite number of solutions all sharing a common conserved value for $T=p_0$ and distinguished only by the value of $x_0$
and thus of $E$ (like there exists an infinite number of solutions all sharing a common conserved value for $E$ and distinguished only by the
couple of values $(x_0,p_0)$ defined by the specific curved $p^2_0/(2m)-x_0F=E$, hence different values for $T=p_0$).

\subsection{Symmetries and their conserved charges}
\label{Sect2.2}

The reason for the existence of the two conserved quantities, $H_0(x,p)$ and $T(p,t)$, of course lies in the existence of symmetries
for the system, namely transformations of its variables $t$ and $x$ such that solutions to the equations of motion are mapped into
one another, and therefore such that these transformations leave invariant these equations of motion.

Of course the existence of the conserved energy of the system, $E=H_0(x,p)$, stems from its invariance under constant translations in time,
$t'=t+t_0$, $x'(t')=x(t)$ and $p'(t')=p(t)$, a set of transformations of parameter $t_0\in\mathbb{R}$ which leave the Lagrange function, $L_0(x,\dot{x})$,
invariant, hence also the equations of motion, since the Lagrange function $L_0(x,\dot{x})$---hence the Hamiltonian $H_0(x,p)$ as well, which is the
conserved generator of this symmetry---is time independent.

Clearly the time dependent conserved quantity $T(p,t)$ must owe its existence to the invariance of the system under constant translations
in space,
\begin{equation}
t'=t,\qquad x'(t')=x(t)\,+\,\xi,\qquad p'(t')=p(t),
\end{equation}
where $\xi\in\mathbb{R}$ is the spatial translation constant parameter. Indeed, this transformation leaves the Lagrangian equation of motion invariant,
whilst in the absence of the external force $F$ it is well known that the momentum conjugate $p$, to which $T(p,t)$ then reduces, is the conserved charge and
generator for such transformations. Nevertheless the Lagrange function $L_0(x,\dot{x})$ is not invariant under such a transformation, since
\begin{equation}
L_0\left(x'(t'),\frac{dx'(t')}{dt'}\right)=L_0\left(x(t),\frac{dx(t)}{dt}\right)\,+\,\xi\,F=L_0\left(x(t),\frac{dx(t)}{dt}\right)\,+\,\frac{d}{dt}\left(\xi\, tF\right).
\end{equation}
However the variation of $L_0$ under the transformation being a surface term, namely a total time derivative, according to
Noether's (first) theorem\cite{JG1} indeed there exists a conserved Noether charge---which is the generator for such transformations for this
system---given by, in the Hamiltonian formulation,
\begin{equation}
T(p,t)=p\,-\,tF.
\end{equation}
But since the quantity of which the total time derivative gives the variation of the Lagrange function is time dependent,
the conserved Noether charge associated to spatial translation invariance carries itself an explicit functional time dependence as an observable over phase space.
Indeed since $T(p,t)$ and $H_0(x,p)$ do not commute in terms of their Poisson bracket in presence of the external force\footnote{Incidentally,
this nonvanishing Poisson bracket defines for the considered system a classical central extension\cite{JLLM1} of the abstract algebra of the Galilei group.},
\begin{equation}
\left\{T(p,t),H_0(x,p)\right\}=F,
\end{equation}
it is the explicit time dependence of the Noether charge $T(p,t)$ which ensures its conservation, given its Hamiltonian equation of motion,
\begin{equation}
\frac{dT(p,t)}{dt}=\frac{\partial T(p,t)}{\partial t}\,+\,\left\{T(p,t),H_0(x,p)\right\}=-F\,+\,F=0,
\end{equation}
while for the Hamiltonian itself one has of course, since it is time independent as a phase space observable,
\begin{equation}
\frac{dH_0(x,p)}{dt}=\left\{H_0(x,p),H_0(x,p)\right\}=0.
\end{equation}

Furthermore, in the same way that the Hamiltonian $H_0(x,p)$ is the generator of infinitesimal constant time translations
given the Hamiltonian equations of motion in phase space, $\dot{x}=\left\{x,H_0(x,p)\right\}$ and $\dot{p}=\left\{p,H_0(x,p)\right\}$,
the time dependent conserved charge $T(p,t)$ is indeed the generator of infinitesimal constant spatial translations since,
\begin{equation}
\left\{x,T(p,t)\right\}=1,\quad \left\{p,T(p,t)\right\}=0,\quad
\delta_\xi x=\left\{x,\xi\,T(p,t)\right\}=\xi,\quad
\delta_\xi p=\left\{p,\xi\,T(p,t)\right\}=0.
\end{equation}

Hence indeed this most simple system possesses two conserved quantities related to its symmetries under constant translations
in time and in space, but such that one of these two infinitesimal generators is time dependent (as a function over phase space)
which then implies that translations in time and in space do not commute with one another (in presence of the external force $F$,
which incidentally includes the case of a constant gravitational acceleration). This is a rather intriguing situation when
considered now within the quantum context and having in mind in particular the gravitational interaction.

\section{When it becomes a Little Quantum Story}
\label{Sect3}

Through the Correspondence Principle\cite{JG1}, let us consider the abstract quantisation of the system
in operator form at the reference time $t=0$. Given the classical phase space observables and their
Poisson brackets, for the corresponding quantum observables at $t=0$ we have the correspondence rule,
\begin{equation}
\left[\hat{x}(0),\hat{p}(0)\right]=i\hbar\,\mathbb{I},\qquad
\hat{x}^\dagger(0)=\hat{x}(0),\qquad
\hat{p}^\dagger(0)=\hat{p}(0),
\label{eq:Heis1}
\end{equation}
while for the conserved Noether charges,
\begin{equation}
\hat{H}_0(0)=\hat{H}_0=\frac{1}{2m}\hat{p}^2(0)\,-\,\hat{x}(0)F,\qquad
\hat{T}(0)=\hat{p}(0),\qquad
\hat{T}(t)=\hat{p}(t)\,-\,tF\,\mathbb{I},
\end{equation}
where in the very last relation $\hat{T}(t)$ stands for the Noether charge for
spatial translations expressed already in the Heisenberg picture of quantum mechanics.

The defining algebra, (\ref{eq:Heis1}), of the quantum system is that of the Heisenberg algebra, of which different
representations are well known\cite{JG1}. Hereafter we shall make use of its configuration and moment
space representations of quantum states. From the outset let us note that we have,
\begin{equation}
\left[\hat{T}(0),\hat{H}_0\right]=i\hbar \,F\, \mathbb{I},\qquad
\left[\hat{T}(t),\hat{H}_0\right]=i\hbar \,F\,\mathbb{I},
\end{equation}
an intriguing fact for these two generators for space and time translations, as already pointed out
in the context of the system's classical dynamics.

\subsection{The Quantum Solution in the Heisenberg Picture}
\label{Subsect3.1}

In the Heisenberg picture, quantum states remain time independent and are considered at the reference time chosen
for quantising the system, $t=0$, while quantum observables, say $\hat{A}(t)$, have a dynamics that evolves according to the
Schr\"odinger equation
\begin{equation}
i\hbar\frac{d\hat{A}(t)}{dt}=i\hbar\frac{\partial\hat{A}(t)}{\partial t}\,+\,\left[\hat{A}(t),\hat{H}_0(t)\right],
\end{equation}
where $\hat{H}_0(t)$ is the Hamiltonian in the Heisenberg picture. In case that $H_0$ does not possess
an explicit time dependence, one has $\hat{H}_0(t)=\hat{H}_0(0)=\hat{H}_0$.

As may easily be checked for the present system, the solutions to this equation for the considered observables read,
\begin{eqnarray}
\hat{x}(t) &=& \frac{1}{2}\frac{F}{m} t^2\mathbb{I}\,+\,\frac{1}{m}\hat{p}(0) t\,+\,\hat{x}(0),\qquad
\hat{p}(t) = \hat{p}(0)\,+\,tF \mathbb{I}, \nonumber \\
\hat{H}_0(t) &=& \frac{1}{2m}\hat{p}^2(0)\,-\,\hat{x}(0) F=\hat{H}_0, \qquad
\hat{T}(t) = \hat{p}(t)\,-\, tF\mathbb{I}=\hat{p}(0)=\hat{T}(0).
\end{eqnarray}
Of course because of the linearity of the equations of motion for these observables, these expressions coincide with the expressions for their classical solutions
with the initial values being replaced by the relevant quantum operators.

\subsection{The Quantum Solution in the Schr\"odinger Picture}
\label{Sect3.2}

In the Schr\"odinger picture quantum states, $|\psi(t)\rangle$, have a dynamics that evolves according to the Schr\"odinger equation,
\begin{equation}
i\hbar\frac{d|\psi(t)\rangle}{dt}=\hat{H}_0\,|\psi(t)\rangle,
\end{equation}
while quantum observables remain considered at the reference time $t=0$,
$\hat{x}\equiv \hat{x}(0)$, $\hat{p}\equiv\hat{p}(0)$, $\hat{H}_0\equiv\hat{H}_0(0)$,
with however the explicit time dependence of the conserved generator for spatial translations
being retained, $\hat{T}(t)=\hat{p}-tF\,\mathbb{I}$.

Since the conserved charges $\hat{H}_0$ and $\hat{T}(t)$ are both hermitian operators, they each possess
a real spectrum of eigenvalues of which the eigenstates span the entire Hilbert space of quantum states and provide
a set of basis vectors in each case. These spectra being continuous and spanning the entire real line, the normalisation of these eigenstates may be
specified\footnote{This choice of normalisation still leaves free to specify an overall constant phase factor for each eigenstate.}
through a Dirac $\delta$-function in their eigenvalues. Namely,
\begin{equation}
\hat{H}_0\,|E\rangle=E\,|E\rangle,\qquad \langle E_1|E_2\rangle = \delta(E_1-E_2),\qquad E\in\mathbb{R},
\end{equation}
\begin{equation}
\hat{T}(t)\,|T,t\rangle=T\,|T,t\rangle,\qquad \langle T_1,t | T_2,t\rangle=\delta(T_1-T_2), \qquad T\in\mathbb{R}.
\end{equation}
Corresponding to these eigenspectra, the solution to the Schr\"odinger equation for the energy eigenstates is of course
\begin{equation}
|E(t)\rangle = e^{-\frac{i}{\hbar}t E}\,|E\rangle,
\end{equation}
while for the $\hat{T}(t)$-eigenstates, $|T,t\rangle$, their complete time dependence is still to be restricted and determined from the Schr\"odinger equation.

As is well known given that the configuration space of the system is simply Euclidean\cite{VV2},
in the configuration space representation using the basis of position eigenstates\footnote{Which is such that
$\hat{x}|x\rangle=x\,|x\rangle$ with $\langle x|x'\rangle=\delta(x-x')$, and thus $\int^{+\infty}_{-\infty}dx\,|x\rangle \langle x|=\mathbb{I}$.},
quantum states are represented through their complex configuration space wave function, $\psi(x,t)=\langle x |\psi(t)\rangle$,
on which the action of the abstract operators $\hat{x}$ and $\hat{p}$ is represented by the functional operators
$x\psi(x,t)$ and $-i\hbar\partial\psi(x,t)/\partial x$, respectively. Likewise in the momentum space representation using the basis of
momentum eigenstates\footnote{Which is such that $\hat{p}|p\rangle=p\,|p\rangle$ with $\langle p|p'\rangle=\delta(p-p')$, 
and thus $\int^{+\infty}_{-\infty}dp\,| p\rangle \langle p|=\mathbb{I}$.}, and with momentum space wave functions
 $\tilde{\psi}(p,t)=\langle p|\psi(t)\rangle$, the same two operators are represented as $i\hbar\partial\tilde{\psi}(p,t)/\partial p$ and $p\tilde{\psi}(p,t)$,
 respectively. Furthermore given that $ \langle x | p \rangle=e^{ixp/\hbar}/\sqrt{2\pi\hbar}$,
 the two types of wave functions are Fourier transforms of one another,
 \begin{equation}
 \psi(x,t)=\int_{-\infty}^{+\infty}\frac{dp}{\sqrt{2\pi\hbar}}\,e^{\frac{i}{\hbar}xp}\,\tilde{\psi}(p,t),\qquad
 \tilde{\psi}(p,t)=\int_{-\infty}^{+\infty}\frac{dx}{\sqrt{2\pi\hbar}}\,e^{-\frac{i}{\hbar}xp}\,\psi(x,t).
 \end{equation} 
 Consequently, in configuration space the two conserved charge operators are,
 \begin{equation}
 \hat{H}_0:\quad -\frac{\hbar^2}{2m}\frac{\partial^2}{\partial x^2}\,-\,x F, \qquad
 \hat{T}(t):\quad -i\hbar\frac{\partial }{\partial x}\,-\, tF,
 \end{equation}
 while in momentum space,
 \begin{equation}
 \hat{H}_0:\quad \frac{1}{2m}p^2\,-\,i\hbar\,F\frac{\partial}{\partial p},\qquad
 \hat{T}(t):\quad p\,-\,tF.
 \end{equation}
 
 Turning first to the energy eigenvalue problem, and denoting
 \begin{eqnarray}
 \psi_E(x,t) &=& \langle x|E(t)\rangle=e^{-\frac{i}{\hbar} tE}\,\langle x|E\rangle=e^{-\frac{i}{\hbar}tE}\,\varphi_E(x), \nonumber \\
 \tilde{\psi}_E(p,t) &=& \langle p|E(t)\rangle=e^{-\frac{i}{\hbar} tE}\,\langle p|E\rangle=e^{-\frac{i}{\hbar}tE}\,\tilde{\varphi}_E(p),
 \end{eqnarray}
 the stationary Schr\"odinger equation becomes,
 \begin{equation}
 \left( -\frac{\hbar^2}{2m}\frac{\partial^2}{\partial x^2}\,-\,x F\right)\varphi_E(x)=E\,\varphi_E(x),\qquad
 \left(\frac{1}{2m}p^2\,-\,i\hbar\,F\frac{\partial}{\partial p}\right)\tilde{\varphi}_E(p)=E\,\tilde{\varphi}_E(p).
 \end{equation}
 Clearly this equation is easier to solve in momentum space. By normalising it as indicated above and with
 a trivial choice of phase factor for $\tilde{\varphi}_E(p=0)$, one finds
 \begin{equation}
 \tilde{\varphi}_E(p)=\frac{1}{\sqrt{2\pi\hbar F}}\,e^{\frac{i}{\hbar}\left(p\frac{E}{F} - \frac{1}{6m}\frac{p^3}{F}\right)},
\qquad E\in\mathbb{R}.
 \end{equation}
 Given the following representation of the Airy function of the first kind\cite{NIST},
 \begin{equation}
 {\rm Ai}\,(x)=\frac{1}{\pi}\,\int_0^\infty\,dt\,\cos\left(\frac{1}{3}t^3\,+\,xt\right),
 \end{equation}
 then the normalised solution in configuration space reads, as is well known,
 \begin{equation}
 \varphi_E(x)=\frac{1}{\sqrt{F}}\,\left(\frac{2mF}{\hbar^2}\right)^{1/3}\,{\rm Ai}\,
 \left(-\left(\frac{2mF}{\hbar^2}\right)^{1/3}\left(x+\frac{E}{F}\right)\right),\qquad E\in\mathbb{R},
 \end{equation}
 with its oscillatory behaviour when $x\ge -E/F$ namely when $V(x)=-xF\le E$, and its exponentially
 decreasing behaviour in the classically forbidden region where $V(x)>E$.

Considering now the eigenvalue problem for the spatial translation operator $\hat{T}(t)$,
\begin{equation}
\left(-i\hbar\frac{\partial }{\partial x}\,-\, tF\right)\psi_T(x,t)=T\,\psi_T(x,t),\qquad
\left(p\,-\,tF\right)\tilde{\psi}_T(p,t)=T\,\tilde{\psi}_T(p,t),
\end{equation}
of which the solution is again easier in momentum space,
\begin{equation}
\tilde{\psi}_T(p,t)=\delta(p - tF - T)\,\tilde{\chi}_T(t)
\end{equation}
with $\tilde{\chi}_T(t)$ some time dependent function still to be determined by imposing now
the Schr\"odinger equation for these quantum states as well as their normalisation as specified above.
Given the choice of a trivial phase factor at $p=0$, one then finds,
\begin{equation}
\tilde{\psi}_T(p,t)=\delta(p - tF - T)\,e^{-\frac{i}{\hbar}\,\frac{t}{2m}(T^2+tF T + \frac{1}{3}t^2F^2)}.
\end{equation}
Consequently in configuration space $\hat{T}(t)$-eigenstates are given by,
\begin{equation}
\psi_T(x,t)=\frac{1}{\sqrt{2\pi\hbar}}\,e^{\frac{i}{\hbar}x\left(T+tF\right)\,-\,\frac{i}{\hbar}\frac{t}{2m}\left(T^2+tF T + \frac{1}{3}t^2 F^2\right)},
\end{equation}
which thus provide another basis for the space of solutions to the time dependent Schr\"odinger equation,
in addition to that of the energy eigenstates, $\psi_E(x,t)=\varphi_E(x)e^{-itE/\hbar}$.

\subsection{Playing around in the Quantum World}
\label{Sec3.3}

First let us consider how the time dependent Noether charge for spatial translations acts on quantum states.
Note that given the above solution for $\hat{T}(t)$-eigenstates, one readily finds,
\begin{equation}
\psi_T(x+\xi,t)=e^{\frac{i}{\hbar}\xi(T+tF)}\,\psi_T(x,t),\quad {\rm namely},\quad
\langle x+\xi | T,t\rangle = e^{\frac{i}{\hbar}\xi(T + tF)}\,\langle x | T,t\rangle,
\label{eq:trans1}
\end{equation}
while on the other hand obviously one also has,
\begin{equation}
\langle x | e^{\frac{i}{\hbar}\xi\hat{T}(t)} | T,t\rangle = e^{\frac{i}{\hbar} \xi T} \, \langle x | T,t \rangle.
\end{equation}
Comparing these two identities, one is led to conclude that the action of the generator of spatial
translations on position eigenstates is such that
\begin{equation}
e^{-\frac{i}{\hbar} \xi \hat{T}(t)}\,|x\rangle = e^{\frac{i}{\hbar} \xi tF}\, |x+\xi\rangle.
\end{equation}
This conclusion may indeed be confirmed based on the completeness relation for the identity
operator in terms of the $\hat{T}(t)$-eigenstates and the observation in (\ref{eq:trans1}),
\begin{eqnarray}
e^{-\frac{i}{\hbar} \xi \hat{T}(t)}\,|x\rangle &=&
\int_{-\infty}^{+\infty}\, dT\,|T,t\rangle \langle T,t|\,e^{-\frac{i}{\hbar} \xi\hat{T}(t)}\,|x\rangle =
 \int_{-\infty}^{+\infty}\,dT\,e^{-\frac{i}{\hbar}\xi T}\,|T,t\rangle \langle T,t|x\rangle \nonumber \\
&=& \int_{-\infty}^{\infty}\,dT\,|T,t\rangle\,e^{\frac{i}{\hbar}\xi tF}\,\langle T,t|x+\xi\rangle =
 e^{\frac{i}{\hbar}\xi tF}\,|x+\xi\rangle.
\end{eqnarray}
Hence even though time dependent the operator $\hat{T}(t)$ is indeed the generator of constant translations
in space. However since it does not commute with the Hamiltonian which itself translates quantum states in time,
as $\hat{T}(t)$ translates a position eigenstate in space
it also changes its phase by a time dependent phase factor, as ought to be the case because of the non vanishing
commutator of $\hat{T}(t)$ and $\hat{H}_0$. On the other hand the spatial translation operator leaves
invariant the momentum eigenstates up to a change in their phase factor as is the case when $F=0$,
but which includes now once again a time dependent phase factor as well,
\begin{equation}
e^{-\frac{i}{\hbar}\xi \hat{T}(t)}\,|p\rangle = e^{-\frac{i}{\hbar} \xi (p-tF)}\,|p\rangle.
\end{equation}
In addition, in spite of its time dependence this Noether generator is indeed conserved as it ought to be,
as shown for instance through the following identities,
\begin{equation}
\hat{T}(t)\,|T,t\rangle = T\,|T,t\rangle,\quad
e^{-\frac{i}{\hbar}\xi \hat{T}(t)}\,|T,t\rangle = e^{-\frac{i}{\hbar} \xi T}\,|T,t\rangle,\quad
\langle T_1,t |\hat{T}(t) | T_2,t\rangle = T_1\,\delta(T_1-T_2).
\end{equation}
Thus $\hat{T}(t)$ is truly the conserved Noether charge for spatial translation invariance of the quantum system,
and yet, it does not commute with its Hamiltonian which is the conserved Noether charge for its time translation invariance.

Given the two bases of eigenstates of this system related to its symmetries, it is also possible to look at this quantum world from complementary perspectives.
Because of the completeness relations for the identity operator in any of these bases,
any quantum state solving the Schr\"odinger equation may be represented in the energy eigenstate basis\footnote{Note that since
$|E(t)\rangle=|E\rangle\,e^{-itE/\hbar}$ while $\langle E(t)|\psi(t)\rangle=\langle E(t=0)|\psi(t=0)\rangle=\langle E|\psi(0)\rangle$, the present
decomposition coincides with the usual expression for it.} as follows,
\begin{equation}
|\psi(t)\rangle =\int_{-\infty}^{+\infty}\,dE\,|E(t)\rangle\,C^E_\psi(E,t),\qquad
C^E_\psi(E,t)=\langle E(t)|\psi(t)\rangle=e^{\frac{i}{\hbar} tE}\,\langle E|\psi(t)\rangle,
\end{equation}
or in the spatial translation eigenstate basis as follows,
\begin{equation}
|\psi(t)\rangle =\int_{-\infty}^{+\infty}\,dT\,|T,t\rangle\,C^T_\psi(T,t),\qquad
C^T_\psi(T,t)=\langle T,t|\psi(t)\rangle.
\end{equation}
In particular it is of interest to consider the distribution of energy eigenstates that builds up any specific
spatial translation eigenstate (or vice-versa). The probability amplitude for this distribution is given by
the matrix elements $C^E_T(E,t)=\langle E(t)|T,t\rangle=e^{itE/\hbar}\,\langle E|T,t\rangle$ which determine the change of basis
in Hilbert space, with
\begin{equation}
\langle E | T,t\rangle = \frac{1}{\sqrt{2\pi\hbar F}}\,
e^{-\frac{i}{\hbar}\frac{t}{2m}\left(T^2+tF T +\frac{1}{3}t^2F^2\right) - \frac{i}{\hbar}\frac{E}{F}\left(T+tF\right) + \frac{i}{\hbar}\frac{1}{6mF}\left(T+tF\right)^3}
=\frac{1}{\sqrt{2\pi\hbar F}}\, e^{\frac{i}{\hbar}\left(\frac{T^3}{6mF}-\frac{E}{F}(T+tF)\right)}.
\end{equation}
Note that since these expressions reduce to pure phase factors, it follows that each of all the possible energy eigenstates contributes
with an equal weight to any of the $\hat{T}(t)$-eigenstates (and vice-versa), since indeed,
\begin{equation}
|C^E_T(E,t)|^2=|\langle E | T,t\rangle)|^2=\frac{1}{2\pi\hbar F}.
\end{equation}
This observation is the quantum analogue of the remark made earlier regarding classical solutions, which are such that
given any prescribed value for $T$ there exists a single solution for whatever value of the energy, $E$. In addition this conclusion also shows
that the quantum states that are eigenstates of spatial translations cannot be energy eigenstates at the same time
since the generators for space and time translations do not commute, while any such state results from a superposition of
all possible energy eigenstates with an equally distributed probability but a specific distribution of time and energy dependent phase factors.
In particular there exists a quantum state truly invariant under all spatial translations, with eigenvalue $T=0$, thus resulting from a specific superposition
of all energy eigenstates. Clearly this is an invitation to explore the possibility of gauging the spatial symmetry for
arbitrary time dependent spatial translations, even though the corresponding conserved Noether charge is time dependent.

\section{Gauging the Story}
\label{Sect4}

Within the classical formulation, since $T(p,t)=p-tF$ is the generator for spatial translations, gauging that symmetry should
be feasible simply by adding to the Hamiltonian of the system an extra term involving a Lagrange multiplier enforcing now
the corresponding constraint over phase space, namely $T(t)=0$. Hence let us consider now the following Hamiltonian first-order
action,
\begin{equation}
S_2=\int\,dt\,L_2,\qquad
L_2=\dot{x}p\,-\,H_2=\dot{x}p\,-\,H_0(x,p)-\lambda(t)(p-tF)
\end{equation}
with $H_2=H_0+\lambda(t) T(p,t)=\frac{1}{2m}p^2-xF+\lambda(t)(p-tF)$, and
where $\lambda(t)\in\mathbb{R}$ is an arbitrary time dependent Lagrange multiplier for the first-class constraint\cite{JG1}
$T(t)=0$. Note that the Hamiltonian of the system, $H_2(x,p,t)$, is thus now explicitly time dependent.

Whether for infinitesimal time dependent spatial translations $\xi(t)$ generated by the constraint,
\begin{equation}
\delta_\xi\lambda(t)=\dot{\xi}(t),\qquad
\delta_\xi x(t)=\left\{x,\xi(t) T(p,t)\right\}=\xi(t),\qquad
\delta_\xi p(t)=\left\{p,\xi(t) T(p,t)\right\}=0,
\end{equation}
or for finite ones,
\begin{equation}
\lambda'(t)=\lambda(t)+\dot{\xi}(t),\qquad
x'(t)=x(t)+\xi(t),\qquad
p'(t)=p(t),
\end{equation}
it may readily be checked that the Hamiltonian first-order Lagrange function transforms only by a surface term,
\begin{equation}
L'_2=L_2\,+\,\frac{d}{dt}\left(\xi(t) t F\right).
\end{equation}
Hence indeed the symmetry under spatial translations has been gauged consistently, even though the corresponding
conserved generator is time dependent.

The equations of motion stemming from the above action now read,
\begin{equation}
\dot{x}=\left\{x,H_2\right\}=\frac{1}{m}p+\lambda,\qquad
\dot{p}=\left\{p,H_2\right\}=F,\qquad
p-tF=0.
\end{equation}
Any solution is still such that $\dot{p}(t)=F$ while however, its acceleration is now totally arbitrary since specified by the
arbitrary Lagrange multiplier, $\ddot{x}(t)=F/m+\dot{\lambda}(t)$, with the following time dependencies,
\begin{equation}
x(t)=x_0+\frac{1}{2}\frac{F}{m}t^2+\int_0^t dt'\lambda(t'),\qquad
p(t)=tF.
\end{equation}
Note that compared to the classical solutions when the symmetry is not gauged,
only one integration constant now is required, $x_0=x(0)$, while that for $p(t)$ is
restricted by the first-class constraint, $p(0)=p_0=0$, since indeed $T(p,t)=p_0=0$.
Furthermore the contribution in $p_0$ to $x(t)$ is now replaced by that which involves
the Lagrange multiplier which itself is not gauge invariant, showing that $x(t)$ is not gauge invariant either.

This observation raises the question of what are the genuine physical, or gauge invariant
degrees of freedom of the system. Clearly the conjugate momentum, $p(t)$, is physical,
but neither are $x(t)$ and $H_0(x,p)$ (with in particular under a gauge transformation, $H'_0=H_0-\xi(t) F$) since,
\begin{equation}
\left\{p,T(p,t)\right\}=0,\qquad
\left\{x,T(p,t)\right\}=1,\qquad
\left\{H_0(x,p),T(p,t)\right\}=-F.
\end{equation}
Consequently besides $p$, the following second quantity is obviously also gauge invariant\footnote{Note that these two gauge
invariant observables commute, $\left\{X(x,p),p\right\}=0$, and are in fact not independent.},
\begin{equation}
X(x,p)=x+\frac{1}{F}H_0(x,p)=\frac{1}{2mF}p^2,\qquad
X(t)=\frac{1}{2m}Ft^2,
\end{equation}
of which the classical solution is indeed independent of the Lagrange multiplier as it should, since
the Lagrange multiplier parametrises the freedom afforded in the choice of degrees of freedom for the
system because of its gauge symmetry\cite{JG1}. However the classical solution for this gauge invariant observable
is not time independent, since it does not commute with the Hamiltonians $H_0$ and $H_2$ of the system,
\begin{equation}
\left\{X,H_0\right\}=\frac{1}{m}p=\left\{X,H_2\right\}=\dot{X}.
\end{equation}

In particular, using the above Hamiltonian equation of motion for $\dot{x}$, it is possible to reduce
the conjugate momentum, $p$, and obtain a Lagrangian action for the gauge system,
\begin{equation}
S_3[x,\lambda]=\int\,dt\,L_3(x,\dot{x},\lambda,t),\qquad
L_3(x,\dot{x},\lambda,t)=\frac{1}{2}m(\dot{x}-\lambda)^2 + xF +t F\lambda.
\end{equation}
It may readily be checked once again that under any gauge transformation,
$t'=t$, $x'(t')=x(t)+\xi(t)$ and $\lambda(t')=\lambda(t)+\dot{\xi}(t)$, the Lagrange function $L_3$
transforms only by a total time derivative as it should. Note that this
Lagrangian formulation of the gauge invariant system now involves an explicitly time dependent
Lagrange function even when considering the Lagrange multiplier as a degree of freedom of 
the dynamics. Finally let us remark that given any possible configuration $x(t)$ of the system,
because of the gauged symmetry under spatial translations it is always possible to gauge that configuration
away to $x'(t)=0$ by applying the gauge transformation with time dependent parameter $\xi(t)=-x(t)$.
Thus any classical solution to the dynamics of the gauged system is gauge equivalent to the configuration
with $x(t)=0$, $p(t)=tF$, hence with the following values for the gauge invariant observables, namely $T(p,t)=0$ as it should while
$X(t)=F t^2/(2m)$.

Turning now to the quantum version of the gauged system, since according to Dirac's quantisation of
constrained systems physical, {\it i.e.}, gauge invariant states
are required\cite{JG1} to be null vectors of the gauge generator $\hat{T}(t)$, thus with vanishing eigenvalue, $T=0$,
in the present instance the physical subspace of Hilbert space reduces to the one-dimensional space
spanned by the $\hat{T}(t)$-eigenstate with eigenvalue $T=0$, whose wave function in configuration space is,
\begin{equation}
\psi_{T=0}(x,t)=\frac{1}{\sqrt{2\pi\hbar}}\,e^{-\frac{i}{\hbar}t\left(\frac{F^2t^2}{6m}\,-xF\right)}.
\end{equation}
Note that the spatial probability distribution of this state is position independent (and identical to that
of the ordinary plane wave solution of the free massive particle),
\begin{equation}
|\psi_{T=0}(x,t)|^2=\frac{1}{2\pi\hbar},
\end{equation}
as is certainly necessary for a state which is to be invariant under arbitrary spatial translations.
We already know that this physical state results from an equally weighted superposition of all possible
energy eigenstates with a specific distribution of time dependent phase factors, since,
\begin{equation}
\langle E | T=0,t\rangle = \frac{1}{\sqrt{2\pi\hbar F}}\,
e^{-\frac{i}{\hbar} t E},\qquad
\langle E(t) | T=0,t\rangle = \frac{1}{\sqrt{2\pi\hbar F}}\,.
\end{equation}
It is remarkable that this result is expressed in such a simple form.

\section{An Invitation as Conclusion}
\label{Sect5}

On first thought it may seem curious that there could exist time dependent conserved quantities.
As this note has shown, this is indeed possible and physically consistent, and arises even in the simplest
example of a nontrivial integrable one degree of freedom system which is that of a nonrelativistic massive particle
subjected to a constant external force, which includes the case of a constant gravitational acceleration.
Time dependent conserved charges are possible but cannot then commute with the Hamiltonian
which itself is the time evolution generator. The linearised explicit time dependence of such Noether charges must then coincide with
their commutator or Poisson bracket with the Hamiltonian. In the case of the system that has been
considered this leads to the situation that translations in time and in space thus do not
commute, a tantalising observation in a gravitational context for which a quantum theory is expected
to have to involve some form of a noncommutative geometry in space(-time).

Furthermore, even though such time dependent conserved charges are generators for time independent
symmetry transformations, such symmetries may be gauged as well and extended to arbitrary time dependent
symmetry transformations. For the simple example having been considered, this note showed such a gauging to
be possible and consistent as well, simply by extending the Hamiltonian formulation in the usual way by adding to
the Hamiltonian, in terms of a set of Lagrange multipliers, a linear combination of the time dependent gauge generators 
now considered as first-class constraints. Since these constraints do not commute with the Hamiltonian,
gauge invariant physical states do not possess a definite energy value, but rather result from
specific superpositions of states of definite energy. In a gravitational context this observation
could indicate that it may be misleading to think in terms of configurations and quantum states of well defined energy.
In the case of the considered example, its symmetry under spatial translations has been gauged
also because gauge invariance under arbitrary space-time diffeomorphisms is a local symmetry which
is central to the relativistic theory of gravity which is General Relativity.

As a matter of fact, the Little Story of this note remains to be completed in a number of different ways,
whose complementary approaches would lead to quantum stories that should need to be compared
and eventually be made physically equivalent. I dare to imagine that Victor would have liked to explore such
paths as well. This is the reason they are offered here as an invitation to further adventures into
the quantum world of the gravitational interaction.

Given a configuration space, it quite obvious that actions which differ only by a surface term, or a total time derivative
in a point particle context, lead to identical equations of motion. Thus by integrating by parts the external force contribution
to the action $S_0$ in (\ref{eq:S0}), an alternative formulation of the dynamics of the same system is provided by the
following action, of which the Lagrange function now displays an explicit time dependence for the velocity coupling to
the constant external force,
\begin{equation}
S_4[x]=\int\,dt\,L_4(x,\dot{x},t),\qquad
L_4(x,\dot{x},t)=\frac{1}{2}m\dot{x}^2\,-\,t\,\dot{x} F.
\end{equation}
The explicit time dependency of the Lagrange function $L_4$ implies that the Hamiltonian of the system in this alternative
formulation is now also explicitly time dependent and thus not conserved for solutions to the classical equations of motion.
And yet, these equations of motion remain certainly invariant under constant translations both in time and in space.
Indeed under constant spatial translations the above action is this time invariant, while rather now it is not invariant
under constant time translations but yet transforms still only by a total time derivative. Consequently, there exist conserved
Noether charges which generate both classes of symmetry transformations. However in the case of the present action, $S_4$,
the generator for spatial translations is now time independent and coincides with the (new) momentum conjugate to the coordinate $x$ which has an expression
in terms of $\dot{x}$ different from that which derives from the Lagrange function $L_0$, while the conserved charge which generates
time translations and defines the conserved energy of the system is now explicitly time dependent as a function over the canonical
phase space that derives from the action $S_4$. Finally, once again the two generators for the symmetries of the
system under translations in space and in time do not commute with one another.

Clearly it would be of interest to apply to the choice of action $S_4$ an analysis totally similar to that of the present note,
and compare results obtained in each approach at every stage of the discussion. In particular, it would be quite interesting to
compare the gauging of the symmetry under spatial translations and the resulting quantum physical state.

Beyond constant translations in space, the dynamics of the model having been considered is also invariant under Galilei boosts,
namely spatial translations which besides constant translations, include also transformations that are linear in time and correspond
to transformations between inertial frames in relative motion at constant velocity. Whether in the case of the action $S_0$ or $S_4$,
an analysis along the lines of the present note may also be developed\footnote{In this instance a classical central extension of the abstract algebra of the Galilei group
arises in the commutator of the two Noether charges generating constant spatial translations and Galilei boosts as is the case already in the
absence of the constant external force\cite{JLLM1}. It may readily be established that the Noether charge for Galilei boosts is given as
$\beta(x,p,t)=mx-tp+Ft^2/2=mx-Ft^2/2-t T(p,t)$, which even though explicitly time dependent, is indeed a third independent conserved quantity for
classical solutions taking then the value $\beta(t)=x_0$. The Poisson bracket of $\beta(x,p,t)$ and $T(p,t)$ determines the classical central extension as
$\left\{\beta(x,p,t),T(p,t)\right\}=m$, while the other classical central extension is $\left\{T(p,t),H_0(x,p)\right\}=F$.}, with in particular the gauging of these symmetry transformations,
since once again the action is invariant under Galilei boosts only up to a total time derivative of a quantity which is explicitly time dependent.
Since gauging these symmetries amounts to requiring that physical configurations are invariant under spatial translations of arbitrary time dependency,
it is to be expected that the resulting quantum physical state ought to coincide in all such analyses.

As mentioned in the Introduction, yet another possible topic of study could be how, given any of the actions $S_0$ or $S_4$,
one has to formulate the same quantum dynamics in arbitrary noninertial frames,
beginning with the frame which is accelerated relative to the considered inertial frame with the constant
acceleration induced by the constant external force, $a=F/m$. Implementing the corresponding spatial translation
which is now quadratically time dependent at the level of the original actions $S_0$ or $S_4$ in order to absorb the contribution
in the external constant force $F$ into the choice of a new coordinate system in the noninertial frame, and then quantising the resulting action,
the question arises as to how the different quantum descriptions are to be compared and possibly reconciled?
What are the roles of the symmetries under time and spatial translations in the new reference frame?
Which quantum state results from gauging once again the symmetry under spatial translations or Galilei boosts in the noninertial frame?
Indeed the time dependent spatial translation absorbing the external force is only one specific  instance of the general
time dependent spatial translations corresponding to the gauged symmetry considered in this note.

The latter study in particular, may also be relevant in understanding better the tension that exists between the local Equivalence
Principle of the gravitational interaction and the basic principles of quantum mechanics. Even though according to Ehrenfest's theorem
the expectation values of the position and momentum quantum observables follow the time dependence of classical trajectories
for the class of models considered in this note, hence these trajectories are also mass independent in the case of the constant gravitational force $F=mg$,
the spreading in time of the wave packet of the quantum particle in free fall should certainly retain a mass dependency (since the mass factors
do not cancel out from the Schr\"odinger equation), thus leading presumably  to observable violations of the gravitational
Equivalence Principle in quantum interference phenomena. By comparing the quantised system either in an arbitrary inertial frame
where this violation would be expected to be observable, or in the noninertial frame which is itself in free fall and such that the gravitational
force would no longer contribute to the Schr\"odinger equation hence presumably leading to a different spreading in time of the wave
function for an identical initial quantum state, could help to better circumscribe what lies at the heart of the tension, or even the fundamental
clash between the local Equivalence Principle of the gravitational interaction and the Principles of Quantum Mechanics, in even as simple a system
as that of the constant gravitational force, $\vec{F}=m\vec{g}$.

\section*{Acknowledgments}

This work is supported in part by the Institut Interuniversitaire des Sciences Nucl\'eaires (I.I.S.N., Belgium),
and by the Belgian Federal Office for Scientific, Technical and Cultural Affairs through the Interuniversity Attraction Poles (IAP) P6/11.

\end{document}